\documentclass[12pt,preprint]{aastex}
\usepackage{graphicx}

\begin{document}
\shortauthors{Pandian et al.}

\title{Physical conditions around 6.7 GHz methanol masers--I: Ammonia}

\author{J. D. Pandian\altaffilmark{1,2}, F. Wyrowski\altaffilmark{3} and K. M. Menten\altaffilmark{3}}
\altaffiltext{1}{Institute for Astronomy, University of Hawaii, 2680 Woodlawn Dr., Honolulu, HI 96822}
\altaffiltext{2}{Indian Institute of Space Science and Technology, Valiamala, Trivandrum 695547, India; jagadheep@iist.ac.in}
\altaffiltext{3}{Max-Planck-Institut f\"{u}r Radioastronomie, Auf dem H\"{u}gel 69, 53121 Bonn, Germany}

\begin{abstract}
Methanol masers at 6.7 GHz are known to be tracers of high-mass star formation in our Galaxy. In this paper, we study the large scale physical conditions in the star forming clumps/cores associated with 6.7~GHz methanol masers using observations of the (1,1), (2,2) and (3,3) inversion transitions of ammonia with the Effelsberg telescope. The gas kinetic temperature is found to be higher than in infrared dark clouds, highlighting the relatively evolved nature of the maser sources. Other than a weak correlation between maser luminosity and the ammonia line width, we do not find any differences between low and high luminosity methanol masers.
\end{abstract}


\section{Introduction}

Molecular masers are frequently used as tools to identify and study regions of star formation. Class~II methanol masers, of which the 6.7~GHz transition is the brightest, are particularly useful as they have been detected only towards high-mass star forming regions. Extensive searches for 6.7~GHz methanol masers towards low-mass star forming regions have not yielded any detections \citep{mini03,bour05,xu08,pand08}. This has been postulated to be due to the lack of a strong far-infrared radiation field in low-mass star forming regions that would be necessary to generate population inversion in regions where the densities are suitable for maser action \citep{pand08}.

One of the questions of interest concerning 6.7~GHz methanol masers is whether or not there is a difference in the properties of sources exciting low and high luminosity masers. The work of \citet{szym00} suggested differences in the colors of IRAS sources associated with weak and strong masers. However, the uncertainty in the positions of both the masers and the IRAS sources was relatively large which raised the question of whether or not the IRAS sources were truly associated with the masers. Furthermore, the later work of \citet{pand07b} did not find any differences between the IRAS colors of weak and strong masers. More recently, using observations of CO and ammonia, \citet{wu10} found several differences between the physical conditions of sources associated with low ($L_{\mathrm{maser}} \sim 10^{-10} - 10^{-7}~L_\sun$) and high luminosity ($L_{\mathrm{maser}} \sim 10^{-6} - 10^{-4}~L_\sun$) 6.7~GHz methanol masers. The primary caveat in this study was the relatively small size of the sample (nine low-luminosity and eight high-luminosity masers). Thus, studies of a larger and more unbiased sample is required to establish the statistical significance of these differences.

The methanol maser catalog of the Arecibo Methanol maser Galactic Plane Survey (AMGPS; \citealt{pand07a}) is ideal for addressing this question. The AMGPS covered the Galactic plane between Galactic longitudes of 35\degr~and 54\degr, and Galactic latitudes, $|b| \lesssim 0.4\degr$. This is the most sensitive blind survey to date for 6.7 GHz methanol masers, and resulted in the detection of 88 sources over 18.2 square degrees. Subsequent follow-up work has determined distances and luminosities of the masers \citep{pand09}. The isotropic maser luminosities calculated from their integrated fluxes range from $4.0 \times 10^{-9}~L_\sun$ to $2.0 \times 10^{-4}~L_\sun$, the median luminosity being $9.1 \times 10^{-7}~L_\sun$. We have been carrying out observations of a number of molecular tracers of dense gas towards the AMGPS sources to study the physical conditions around 6.7 GHz methanol masers. In this paper, we report results from observations of the $(J,K)$ = (1,1), (2,2) and (3,3) inversion transitions of ammonia using the Effelsberg 100~m telescope\footnote{Based on observations with the 100-m telescope of the MPIfR (Max-Planck-Institut f\"{u}r Radioastronomie) at Effelsberg.}.

The inversion transitions of ammonia are a powerful tool to measure the physical conditions in star forming regions \citep{ho83}. The detection of the hyperfine structure of the inversion lines allows one to measure their optical depth. Since radiative transitions between the different $K$-ladders of the $(J,K)$ levels are forbidden, one can obtain a calibration independent estimate of the gas temperature using observations of the (1,1) and (2,2) lines. Moreover, in contrast to many carbon bearing molecules, ammonia is resistant to depletion onto dust grains (e.g. \citealt{berg97}). Thus, ammonia is an excellent tracer of the dense gas in star forming regions. Here, we use ammonia observations towards the AMGPS masers in combination with complementary 1.1~mm continuum data from the Bolocam Galactic Plane Survey (BGPS; \citealt{agui11}) to determine temperatures, column densities and ammonia abundances, and examine if these properties differ between low and high luminosity methanol masers.

\section{Observations and data reduction}

The observations were carried out between 2009 April 30 and May 7 using the Effelsberg 100~m telescope. We used the cooled K-band HEMT receiver to observe two orthogonal linear polarizations. The backend used was a modified version of the fast Fourier transform spectrometer (FFTS) used at the APEX telescope \citep{klei06}. The FFTS was used with a bandwidth of 500 MHz split into 16,384 channels giving a velocity resolution of 0.4~km~s$^{-1}$ at the frequency of the NH$_3$ (1,1) inversion transition. The receiver was tuned to a frequency of 23.78 GHz allowing simultaneous observation of the (1,1), (2,2) and (3,3) transitions. The observations were made in frequency switched mode with a frequency throw of 7.5~MHz. The integration time ranged from 3 to 15 minutes per source depending on the strengths of the lines. Pointing and focus checks were carried out periodically, with the pointing accuracy being found to be better than 10$''$. The flux density scale was set using observations of NGC~7027 which was assumed to have a flux density of 5.44 Jy at the observed frequency \citep{ott94}. This in turn was converted to main beam brightness temperature ($T_{\mathrm{MB}}$) assuming a conversion factor of 1.36~K~Jy$^{-1}$ (using eq. 8.20 of Rohlfs \& Wilson 2004). The flux scale is estimated to be accurate to $\sim 10\%$. The half-power beam width of the telescope at the observed frequency is $\sim 40\arcsec$.

The data were reduced using the CLASS software package of the GILDAS distribution\footnote{http://www.iram.fr/IRAMFR/GILDAS}. A polynomial baseline was subtracted followed by folding of the frequency switched spectra. 

\section{Results and analysis}

\begin{figure}[!htb]
\centering
\includegraphics[height=0.9\textheight]{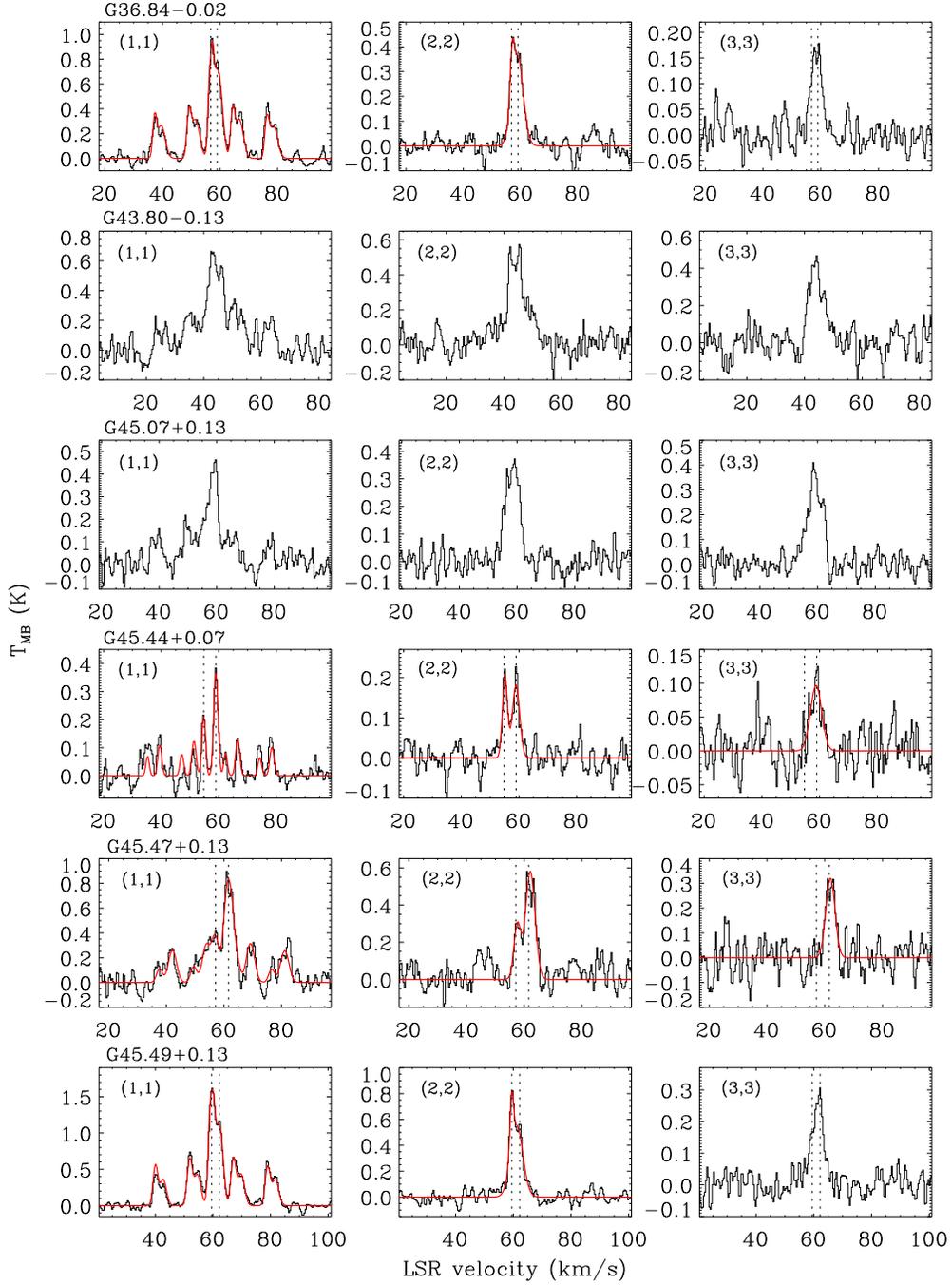}
\caption{NH$_3$ (1,1), (2,2) and (3,3) spectra for sources with multiple lines. The red curve shows the spectral line fit. Spectra with no fits cannot be fit reliably with consistent velocities in all three transitions.}\label{complexspec}
\end{figure}

The ammonia observations were carried out towards 77 out of 88 AMGPS sources. The W49N and W51main sources were excluded due to the crowded fields which would result in multiple sources within the Effelsberg beam. G41.87$-$0.10 was excluded due to the uncertainty in its position, while G49.42$+$0.32 is included in the observation towards G41.41$+$0.33 on account of the two sources being only $\sim 2''$ apart. The NH$_3$ (1,1) inversion transition was detected in 73 out of 77 sources. The (2,2) and (3,3) transitions were detected in 63 and 43 sources respectively. Six sources displayed spectra characteristic of multiple lines or complex line profiles (Figure~\ref{complexspec}).

The NH$_3$ (1,1) spectra were fit using the standard NH$_3$ (1,1) method in CLASS, which takes into account the hyperfine structure and line broadening effects. Some sources were too weak to fit the hyperfine components. In these cases, we fit the (1,1) spectra using Gaussians. We fit the (2,2) and (3,3) spectra using Gaussians since the hyperfine structure in these lines was not detected in most cases. The results of the line fitting are tabulated in Table 1. Of the six sources with more than one peak, two could not be fit reliably using multiple lines (see Figure~\ref{complexspec}). These sources are excluded in Table~1. The remaining four sources could be fit using two velocity components in both (1,1) and (2,2) transitions, and the two components are distinguished by the suffix `a' and `b' in Table~1. For two out of these four sources, the two velocity components could not be fit for the (3,3) line, which is typically broader than the (1,1) and (2,2) lines (and is hence more blended). These are noted in Table~1.

\subsection{Determining physical parameters}

The observation of multiple transitions allows one to analytically determine physical parameters such as temperature, density and column density in the sources. The methodology and the equations used are discussed in Appendix~\ref{app1}.

The use of eq.~(\ref{texeq}) requires knowledge of the optical depth of the line. As indicated earlier, some sources were too weak to fit the hyperfine components (G37.38$-$0.09, G37.53$-$0.11 and G42.43$-$0.26) and consequently, their optical depths of the (1,1) main hyperfine component are unknown. For these sources, $T_{\mathrm{ex}}$ is assumed to be 4.1~K (which is the mean excitation temperature of the other sources).

In determining the kinetic temperatures from eq.~(\ref{tkineq}), we used collision coefficients from \citet{danb88} who provide the coefficients for different values of $T_\mathrm{k}$. We iterated the procedure until the value adopted for $T_\mathrm{k}$ (used to obtain collision coefficients) was consistent with that derived from eq.~(\ref{tkineq}).

The derived values of the physical parameters of the sources are listed in Table 2. Only sources with at least $4\sigma$ detections of at least two NH$_3$ transitions are shown in Table~2. In all cases, uncertainties in the physical parameters are derived by propagating the uncertainties in the variables (the 10\% uncertainty in flux calibration is added in quadrature to the formal uncertainty of the main beam antenna temperatures in Table~1) of the relevant equations -- i.e. if $y = f(x_1,x_2,\ldots,x_n)$, then
\begin{equation}
\sigma_y^2 = \sum_{i=1}^n~\left|\frac{\partial f}{\partial x_i}\right|^2~\sigma_{x_i}^2
\end{equation}
assuming that the different variables are uncorrelated (which is not strictly true, but is a good approximation for our purpose). The exception is for kinetic temperatures corresponding to $T_{\mathrm{R21}}$ greater than $\sim 25$~K since the behavior of $T_\mathrm{k}$ becomes highly non-linear at that point (see Figure~2a of \citealt{walm83}). For these cases, the uncertainties in $T_\mathrm{k}$ are calculated by estimating the kinetic temperature corresponding to $T_{\mathrm{R21}} \pm \sigma_{T_{\mathrm{R21}}}$ in eq.~(\ref{tkineq}).  The uncertainties in the physical parameters are typically dominated by the uncertainty in the optical depth of the (1,1) main hyperfine component. 

Note that the excitation temperatures are typically found to be a factor of 4 lower than the rotation temperature $T_{\mathrm{R21}}$. This could be due to either sub-thermal excitation of the ammonia lines, or due to a small beam filling factor of the emitting gas. Considering that all sources are expected to be in regions of high-mass star formation (and are quite distant with distances greater than 3.5~kpc), it is much more likely that small beam filling factors are responsible for the low excitation temperatures. Hence, the column densities shown in Table~2 have been derived using $T_{\mathrm{R21}}$ rather than $T_{\mathrm{ex}}$ (which is derived from eq.~(\ref{texeq}) assuming $\eta_b$ to be unity).

\section{Discussion}

In discussing the physical conditions of the sources, it should be kept in mind that the half power beamwidth of the telescope is $\sim 40''$ at the wavelength of the (1,1), (2,2) and (3,3) lines. Hence, the derived physical conditions are characteristic of the large scale conditions in the star forming clump/core, and are not indicative of the conditions in the masing clumps themselves, which have sizes that are smaller by one to two orders of magnitude.

\subsection{Temperature}
\begin{figure}[!htb]
\centering
\includegraphics[width=0.45\textwidth]{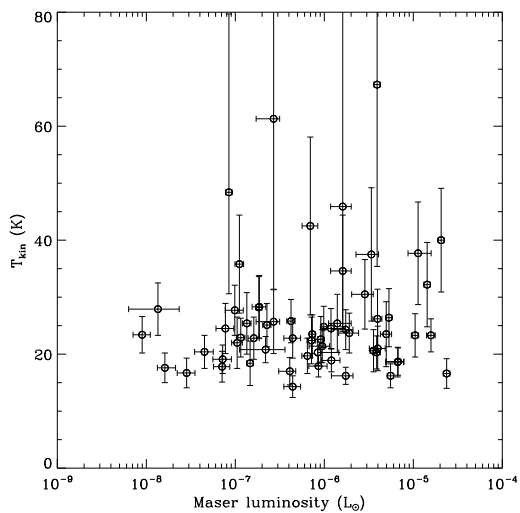}
\caption{NH$_3$ kinetic temperature as a function of maser luminosity.}\label{tkinfig}
\end{figure}

The mean and median values of the kinetic temperature of the sources are 26.0~K and 23.4~K respectively. The kinetic temperatures are slightly lower (though comparable when considering the standard deviation) than that of the methanol maser sample of \citet{wu10} (mean kinetic temperature of 33.3~K). In contrast, the mean kinetic temperature of a sample of infrared dark clouds studied by \citet{pill06} is 14.8~K. This is most likely to be an evolutionary trend since methanol maser sources are more evolved compared to infrared dark clouds. A small number of sources have had their spectral energy distributions modeled resulting in estimates of the temperature of the central star \citep{pand10}. Although the number of sources is too small to be statistically significant, we find that sources with higher stellar temperatures also tend to have higher ammonia kinetic temperature, which bolsters the suggestion of higher kinetic temperatures being an evolutionary trend. There is no correlation between the luminosity of methanol masers and the kinetic temperature (Figure~\ref{tkinfig}), which is consistent with previous results \citep{wu10}. We also do not see any correlation between the kinetic temperature and other maser properties such as its overall velocity extent.

\begin{figure}[!htb]
\centering
\includegraphics[width=\textwidth]{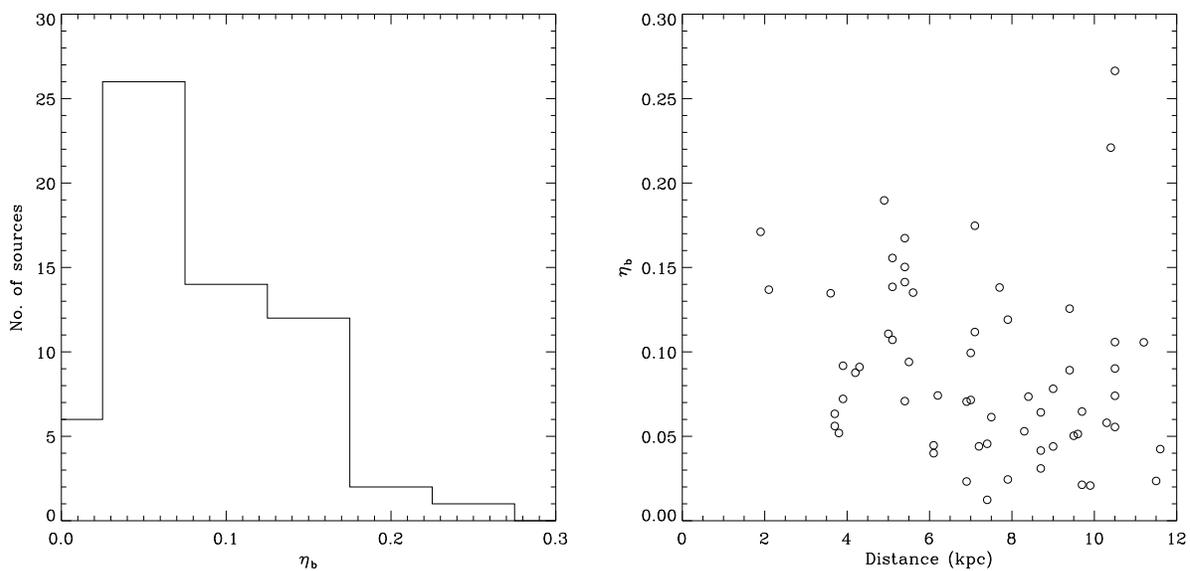}
\caption{The left panel shows the histogram of the beam dilution factor, $\eta_b$, calculated assuming the excitation temperature of the (1,1) line to be equal to the rotation temperature, $T_{\mathrm{R21}}$. The right panel shows $\eta_b$ as a function of the distance to the source.}\label{bdilfig}
\end{figure}

The excitation temperature of the (1,1) line ranges from 3.0~K to 7.4~K. The mean value of $T_{\mathrm{ex}}$ is 4.1~K, the standard deviation being 0.9~K. These values are much lower than the rotation temperature determined from the (1,1) and (2,2) lines, which is most likely to be due to the small beam filling factor of the dense gas.
Assuming that the excitation temperature is equal to the rotation temperature ($T_{\mathrm{R21}}$), one can determine the beam filling factor of the emitting gas from eq.~(\ref{texeq}). This in turn can be used to estimate the physical sizes of the emitting regions since the distances to the sources are known. We find the beam filling factor to have a median value of 0.07. The average physical size of the emitting region is found to be 0.1~pc, which is typical for a massive star forming core (e.g. \citealt{mott07}). The left panel of Figure~\ref{bdilfig} shows the distribution of the beam filling factor in our sources. The distribution is similar to that seen in massive dense cores selected from the ATLASGAL survey (Wienen et al., in prep). The right panel of Figure~\ref{bdilfig} shows the beam filling factor as a function of distance to the source. If our hypothesis of low excitation temperatures being indicative of small beam filling factors is true, the deduced beam filling factor should decrease with the distance to the source. This is indeed found to be the case, with the correlation coefficient between the beam filling factor and distance being $-$0.45 (ignoring the two outliers at large distances). In light of this discussion, it is clear that the differences in the excitation temperature between low and high luminosity methanol masers found by \citet{wu10} is more likely to be due to differences in the distances to the two samples and the resulting differences in the beam filling factor rather than a real physical effect. The mean distance to the low-luminosity maser sample of \citet{wu10} is 1.7~kpc, while that of the high-luminosity sources is more than a factor of two higher at 3.6~kpc. This would also explain the much lower excitation temperatures seen in our sample since the AMGPS sources are much more distant with a mean distance of 7.2~kpc.

\begin{figure}[!htb]
\centering
\includegraphics[width=0.45\textwidth]{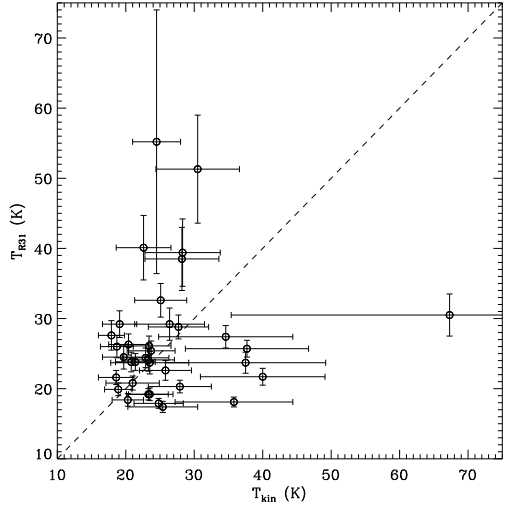}
\caption{The rotation temperature determined from (3,3) and (1,1) transitions as a function of the kinetic temperature. The dashed line shows where $T_{\mathrm{R31}}$ is equal to $T_\mathrm{k}$.}\label{trotfig}
\end{figure}

As can be seen from Table 2, the rotation temperature determined from the (1,1) and (3,3) lines ($T_{\mathrm{R31}}$) is generally larger than that determined from the (1,1) and (2,2) lines ($T_{\mathrm{R21}}$). This is most likely due to the presence of temperature gradients in the sources with the (3,3) line tracing warmer gas in the star forming clumps. It is however interesting to see that $T_{\mathrm{R31}}$ is much larger than the kinetic temperature ($T_\mathrm{k}$) in a few cases (Figure~\ref{trotfig}). This could potentially be due to some contribution to the (3,3) intensity from population inversion \citep{walm83}. However, since the (1,1) and (3,3) lines arise from two different species (para versus ortho-NH$_3$), some variation could also be explained from different ortho to para ratios in different sources.

\subsection{Line widths}

The NH$_3$ (1,1) line widths range from 1.3~km~s$^{-1}$ to 4.1~km~s$^{-1}$, the mean line width being 2.2~km~s$^{-1}$. This is comparable with previous studies of infrared dark clouds \citep{pill06} and 6.7 GHz methanol masers \citep{wu10}. As is generally found in star forming regions, the line widths are much larger than the thermal line widths (which is $\sim 0.26$~km~s$^{-1}$ at a temperature of 25~K). This is generally interpreted as being due to turbulence and clumping within the single dish beam. The (2,2) line width ranges from 1.4~km~s$^{-1}$ to 5.2~km~s$^{-1}$, which is larger than the line width of the (1,1) line. The mean ratio of the (2,2) to (1,1) line widths is 1.3. Similarly, the (3,3) line displays a larger line width ranging from 2.1~km~s$^{-1}$ to 5.6~km~s$^{-1}$, the mean ratio of the (3,3) to (1,1) line widths being 1.6. 

The larger line widths of the (2,2) and (3,3) lines are in part due to optical depth effects that are not taken into account by the Gaussian fitting. For example, using sensitive observations that detect the hyperfine structure in the (2,2) line, Wienen et al. (in prep.) find that the ratio of the apparent (2,2) line width (based on Gaussian fitting) to that of the (1,1) line width is $\sim 1.15$. This however will still not explain the observed increase in the line widths of the (2,2) and (3,3) lines. These observations are consistent with previous studies. The higher line widths of the higher $J$ transitions is most likely due to two reasons. The higher $J$ transitions are likely to arise from warmer gas on account of the higher energy levels involved. The warmer gas is likely to display a more clumpy structure with differential motions within the single dish beam compared to the cooler envelope that is traced by the lower $J$ transitions. However, a similar observation is seen even in the high angular resolution study of \citet{long07}, who suggest systemic motions such as rotation, infall or outflow, or turbulence injection to be the source of larger line widths in higher $J$ transitions.

\begin{figure}[!htb]
\centering
\includegraphics[width=\textwidth]{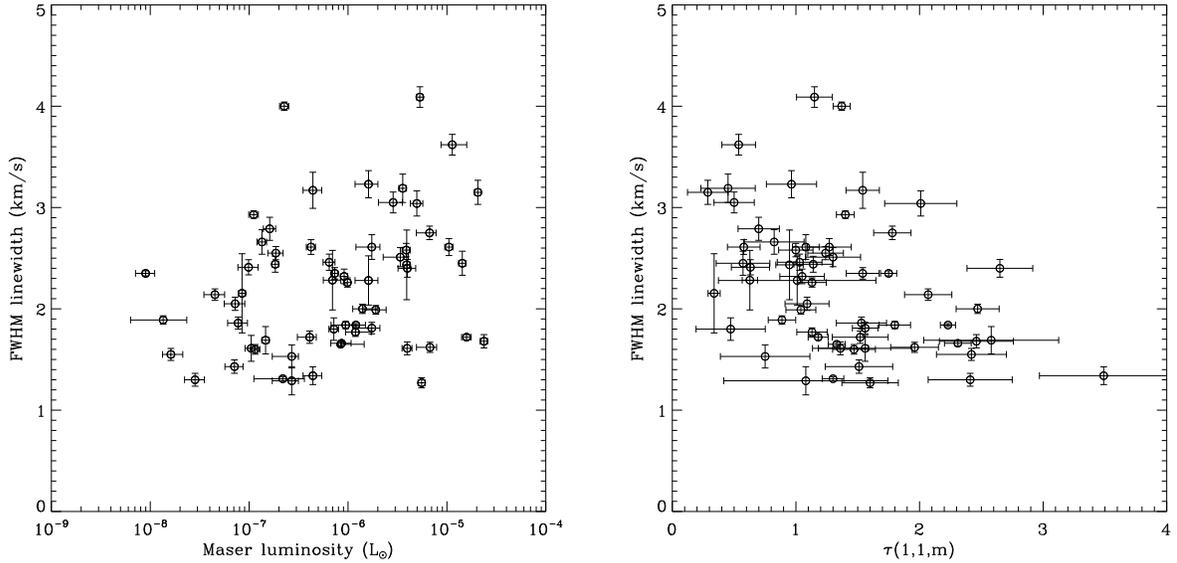}
\caption{NH$_3$ (1,1) line width as a function of maser luminosity (left panel) and the optical depth of the (1,1) main hyperfine component (right panel).}\label{vwidthfig}
\end{figure}

The left panel of Figure~\ref{vwidthfig} shows the distribution of the (1,1) line width as a function of maser luminosity. The previous study of \citet{wu10} noted a strong correlation between the maser luminosity and the ammonia line width (correlation coefficient of 0.8). In contrast, we see only a weak correlation (correlation coefficient of 0.3, improving to 0.45 if five data points in the lower right are not considered) in our data. This discrepancy is most likely to be due to the relatively small sample size of \citet{wu10}. We also see an anti-correlation between the line width and the optical depth of the (1,1) line (correlation coefficient of $-$0.4; Figure~\ref{vwidthfig}, right panel), similar to what was noted by \citet{long07}. Though interesting, the physical reason behind such a correlation is not clear.

\subsection{Column densities and Abundances}

\begin{figure}[!htb]
\centering
\includegraphics[width=0.45\textwidth]{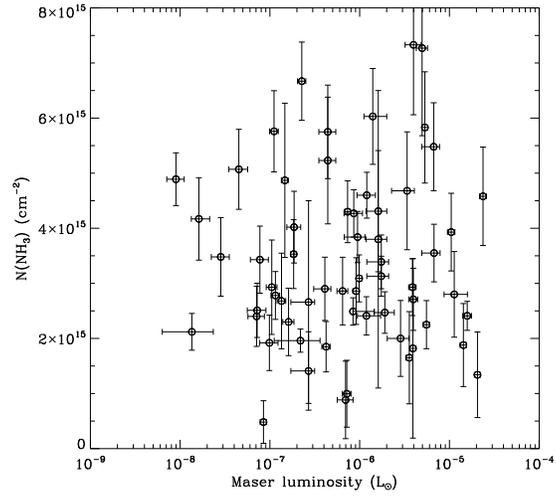}
\caption{NH$_3$ column density as a function of maser luminosity.}\label{densfig}
\end{figure}

The NH$_3$ column density ranges from $4.8 \times 10^{14}$~cm$^{-2}$ to $7.3 \times 10^{15}$~cm$^{-2}$, the median value being $2.9 \times 10^{15}$~cm$^{-2}$. This is comparable to the column densities seen in infrared dark clouds \citep{pill06}, and higher than the column densities of low-luminosity methanol masers \citep{wu10}. The latter could be due to our assumption of LTE conditions in our sources. If the maser sources of \citet{wu10} are also assumed to be in LTE, the resulting NH$_3$ column densities would be a factor of 2$-$3 higher, which would be similar to what is seen in our sample. Figure~\ref{densfig} shows the NH$_3$ column density as a function of luminosity. There is clearly no correlation between the maser luminosity and NH$_3$ column density. This is very similar to the relation seen in \citet{wu10} for maser luminosities $\lesssim 10^{-5}~L_\odot$ (see right panel of Figure~7 in \citealt{wu10}). However, \citet{wu10} see a sharp rise in the NH$_3$ column density at higher maser luminosities. This cannot be verified in this study since the highest maser luminosity in our sample for which unambiguous line fits could be performed is only $2.4 \times 10^{-5}~L_\odot$.

The volume density of H$_2$ molecules can be estimated from the $(1,1)$ line using eq.~(2) of \citet{ho83}:
\begin{equation}\label{nh2eq}
n(\mathrm{H}_2) = \frac{A}{C}\left[\frac{J_\nu(T_{\mathrm{ex}})-J_\nu(T_{\mathrm{bg}})}{J_\nu(T_{\mathrm{k}})-J_\nu(T_{\mathrm{ex}})}\right]\left[1 + \frac{J_\nu(T_\mathrm{k})}{h\nu/k}\right]
\end{equation}
where $A$ is the Einstein $A$ coefficient ($=1.71 \times 10^{-7}$~s$^{-1}$) and $C$ is the collisional de-excitation rate ($C(1,1,a;1,1,s)$ in \citealt{danb88}). However, beam dilution will result in a significant underestimation of the densities. The assumption of LTE would cause eq.~(\ref{nh2eq}) to diverge. We hence estimate the column density of H$_2$ using the BGPS catalog \citep{roso10}. For each maser source, we searched for 1.1~mm counterparts within a search radius of 20$''$ (Table 3). The H$_2$ column density, $N$(H$_2$) can be calculated from the observed flux density, $S_\nu$ using
\begin{equation}\label{h2coldens}
S_\nu = \Omega B_\nu(T_d) \kappa_\nu \mu m_H N(\mathrm{H}_2)
\end{equation}
where $\Omega$ is the solid angle of the source, $B_\nu(T_d)$ is the black body function at dust temperature, $T_d$, $\kappa_\nu$ is the dust opacity, $\mu$ is the mean molecular weight assumed to be 2.8, and $m_H$ is the mass of the hydrogen atom. The source solid angle is calculated from the its angular size listed in the BGPS catalog. We adopt a dust opacity of 0.01~cm$^2$~g$^{-1}$ at 1.1~mm from \citet{osse94} assuming a gas to dust ratio of 100, and a dust temperature equal to the gas kinetic temperature (derived from the NH$_3$ data). In calculating H$_2$ column densities, the fluxes listed in the BGPS catalog (and shown in Table~3) have been multiplied by a factor of 1.5, which is required to make the fluxes consistent with those of other published surveys \citep{agui11}.

The H$_2$ column densities for sources that also have measured NH$_3$ column densities are listed in Table 3, and range from $1.2 \times 10^{21}$~cm$^{-2}$ to $2.2 \times 10^{22}$~cm$^{-2}$ with a median value of $7.4 \times 10^{21}$~cm$^{-2}$. The ammonia abundance can be estimated from the H$_2$ and NH$_3$ column densities as
\begin{equation}
\chi({\mathrm{NH}}_3) = \frac{N(\mathrm{NH}_3)}{N(\mathrm{H}_2)}
\end{equation}
The NH$_3$ abundance ranges from $7.0 \times 10^{-8}$ to $7.4 \times 10^{-7}$ with a median value of $2.7 \times 10^{-7}$. This is almost an order of magnitude higher than typical abundances derived for star forming regions (e.g. \citealt{bens83,harj93,berg97,lang00,pill06}). However, this is very similar to the abundances derived for massive cores selected from the ATLASGAL survey (Wienen et al., in prep). One of the possible reasons for the high ammonia abundance may be that the poor resolution of the BGPS maps would result in the H$_2$ column densities being underestimated. We observe a slight anti-correlation between the H$_2$ column density and the distance to the source, which is expected if the cores are unresolved. Assuming the same beam filling factor as for the ammonia emission would reduce the solid angle of the cores by a factor of $\sim 14$. Taking into account that individual cores would have a lower integrated flux than that of the large scale clump measured from the BGPS, the H$_2$ column density would increase roughly by an order of magnitude. This would reduce the ammonia abundance by an order of magnitude bringing it much closer to the values in the literature. Observations at higher angular resolution that resolve the cores from the larger (and less dense) clumps can verify this hypothesis.



\subsection{Are there distinctions between low and high luminosity methanol masers?}

The work of \citet{wu10} found a number of distinctions between low and high luminosity 6.7 GHz methanol masers. In particular, strong correlations were found between the line widths and luminosity, and the cores associated with more luminous masers were found to be more massive. An increase in the ammonia column density was also found for the high luminosity masers, leading to suggestions that more luminous masers were associated with higher mass stars. An alternate possibility suggested by \citet{wu10} is that higher luminosity masers were associated more evolved sources, which was bolstered by the work of \citet{bree10} who found that more luminous 6.7 GHz masers were more likely to be associated with OH masers (which tend to be associated with more evolved sources).

The work presented in this paper, comprising of a larger sample of masers, does not find much distinction between low and high luminosity methanol masers. We find a weak correlation between the maser luminosity and ammonia line width, and no correlation between the maser luminosity and the other physical quantities. The high angular resolution observations of \citet{long07} found that the ammonia line width was correlated with the evolutionary state of the cores, as determined from the presence or absence of 6.7 GHz methanol masers and 24~GHz continuum emission. Although this cannot be applied directly to our current study due to our much poorer angular resolution (8\arcsec~of \citealt{long07} versus 40\arcsec~in this work), the weak correlation that we observe between maser luminosity and ammonia line width suggests that there is no strong correlation between maser luminosity and the evolutionary state of the source. Moreover, since the ammonia kinetic temperature appears to be correlated with the evolutionary stage of a source, one would expect to see a correlation between the kinetic temperature and maser luminosity if luminous masers are indeed associated with more evolved sources. The lack of any such correlation again casts some doubt on this hypothesis of \citet{wu10}. 

Since the existing millimeter data is at relatively poor resolution which results in a systematic bias towards higher masses at larger distances, we do not look for any correlation between the maser luminosity and the mass of the core exciting the maser. However, it is to be noted that there is a difference of a few orders of magnitude between the size scale of a masing clump and that of a star forming core. Moreover, maser radiation involves exponential amplification of seed radiation and is dependent on the geometry of the source. Hence, it is not clear whether one should expect any correlation between the maser properties, and the large scale properties of the star forming core/clump. It may be possible to see correlations if the observations had sufficient angular resolution to probe individual masing clumps. However, given the poor brightness sensitivities of interferometers at such small scales, such an experiment may not be feasible.

\section{Conclusions}

We have observed the (1,1), (2,2) and (3,3) inversion transitions of ammonia towards the 6.7 GHz methanol masers detected in the AMGPS. The excitation temperature of the (1,1) line is seen to be much smaller than the rotation temperature determined from the (1,1) and (2,2) lines, suggestive of small beam filling factors for the ammonia emission. The kinetic temperatures are warmer than those seen in infrared dark clouds, suggestive of an evolutionary trend towards warmer temperatures with the age of the source. We see only a weak correlation between the maser luminosity and the line width of the ammonia lines. No other correlations are seen between the properties of the methanol masers and those of the star forming clumps/cores. While this may be in part due to the relatively poor resolution of the current observations, it does cast some doubt on whether there are distinctions between low and high luminosity methanol masers.

\begin{appendix}
\section{Determining physical parameters from ammonia}\label{app1}

In the following, the methodology used for calculating excitation temperature, rotation temperature, kinetic temperature, and column density from observations of the (1,1), (2,2), and (3,3) inversion transition of NH$_3$ is discussed.

The excitation temperature ($T_{\mathrm{ex}}$) is determined from the observed main beam antenna temperature ($T_{\mathrm{MB}}$) using 
\begin{equation}\label{texeq}
T_{\mathrm{MB}} = \eta_b\left[J_\nu(T_{\mathrm{ex}}) - J_\nu(T_{\mathrm{bg}})\right](1-e^{-\tau})
\end{equation}
where $\eta_b$ is the beam filling factor, $T_{\mathrm{bg}}$ is the background temperature (assumed to be 2.73 K), $\tau$ is the optical depth of the line, and $J_\nu(T) = (h\nu/k)\left[\exp(h\nu/kT)-1\right]^{-1}$. 

The rotation temperature can be calculated from observations of two transitions, $(J,K)$ and $(J',K')$ using eq. (3) of \citet{ho83}
\begin{equation}\label{tauratio1eq}
\frac{\tau(J',K')}{\tau(J,K)} = \frac{\nu^2_{J',K'}}{\nu^2_{J,K}}\frac{\Delta\nu_{J,K}}{\Delta\nu_{J',K'}}\frac{T_{\mathrm{ex}}(J,K)}{T_{\mathrm{ex}}(J',K')}\frac{|\mu_{J',K'}|^2}{|\mu_{J,K}|^2}\frac{g_{J,'K'}}{g_{J,K}}\exp\left(-\frac{\Delta E_{J'K',JK}}{kT_\mathrm{R}(J',K';J,K)}\right)
\end{equation}
where $\nu_{J,K}$ and $\Delta\nu_{J,K}$ are the frequency and line width of the $(J,K)$ transition respectively, $\mu_{J,K}$ is the dipole matrix element given by $|\mu_{J,K}|^2 = \mu^2 K^2/[J(J+1)]$ ($\mu = 1.468$ Debye), $g$ is the statistical weight, and $\Delta E$ is the energy difference between the two states, and $T_\mathrm{R}$ is the rotation temperature. For the (1,1), (2,2) and (3,3) lines, it is common to assume that they have the same line widths (although see the discussion in Sect. 4) and excitation temperatures. In most cases, the optical depth is high enough to be measurable only for the main hyperfine component of the (1,1) transition. The optical depth of the other transitions can be estimated from that of the (1,1) transition using eq. (A3) of \citet{mang92}
\begin{equation}\label{tauratio2eq}
\frac{T_{\mathrm{MB}}(J,K,m)}{T_{\mathrm{MB}}(J',K',m)} = \frac{1-\exp[-\tau(J,K,m)]}{1-\exp[-\tau(J',K',m)]}
\end{equation}
under the assumption that the beam filling factors and excitation temperatures of the different transitions are the same, and that the transitions are in local thermodynamic equilibrium. The optical depth of a $(J,K)$ transition is related to that of the main hyperfine transition by a simple scale factor:
\begin{equation}\label{tautranseq}
\tau(J,K) = a\tau(J,K,m)
\end{equation}
where $a$ is 2.0 for the (1,1) transition, 1.256 for the (2,2) transition and $\sim 1.124$ for the (3,3) transition. Using eqs.~(\ref{tauratio1eq}) to (\ref{tautranseq}) above, one can derive the rotation temperature between the (1,1) and (2,2) transitions ($T_{\mathrm{R21}}$) as
\begin{equation}
T_{\mathrm{R21}} = -41.2\div \ln\left\{-\frac{0.283}{\tau(1,1,m)}\ln\left[1-\frac{T_{\mathrm{MB}}(2,2,m)}{T_{\mathrm{MB}}(1,1,m)}\left(1-e^{-\tau(1,1,m)}\right)\right]\right\}
\end{equation}
Similarly, the rotation temperature between the (1,1) and (3,3) transitions ($T_{\mathrm{R31}}$) can be written as
\begin{equation}
T_{\mathrm{R31}} = -100.3\div \ln\left\{-\frac{0.0803}{\tau(1,1,m)}\ln\left[1-\frac{T_{\mathrm{MB}}(3,3,m)}{T_{\mathrm{MB}}(1,1,m)}\left(1-e^{-\tau(1,1,m)}\right)\right]\right\}
\end{equation}

To determine the kinetic temperature ($T_{\mathrm{k}}$) from $T_{\mathrm{R21}}$, we follow the approach of \citet{walm83} and assume that the (1,1), (2,2) and (2,1) levels form a three level system. Then, if $n_{1,1}$ and $n_{2,2}$ are the populations of the (1,1) and (2,2) levels, the collisional balance in the three level system can be written as
\begin{equation}
n_{2,2}[C(2,2;1,1) + C(2,2;2,1)] = n_{1,1}C(1,1;2,2)
\end{equation}
where $C(J,K;J',K')$ is the collisional excitation/de-excitation rate from the $(J,K)$ level to the $(J',K')$ level. Note that $C(J,K;J',K')$ is related to $C(J',K';J,K)$ by the relation
\begin{equation}
g_{J,K}C(J,K;J',K') = g_{J',K'}C(J',K';J,K)~e^{-\Delta E_{J'K',JK}/kT_{\mathrm{k}}}
\end{equation}
Moreover, $n_{1,1}$ and $n_{2,2}$ are related by
\begin{equation}
\frac{n_{2,2}}{n_{1,1}} = \frac{5}{3}~e^{-41.2/T_{\mathrm{R21}}}
\end{equation}
Hence, the kinetic temperature is related to $T_{\mathrm{R21}}$ by
\begin{equation}\label{tkineq}
1+\frac{C(2,1;2,2)}{C(2,2;1,1)}\exp\left(-\frac{16.0}{T_{\mathrm{k}}}\right) = \exp\left[41.2\left(\frac{1}{T_{\mathrm{R21}}} - \frac{1}{T_\mathrm{k}}\right)\right]
\end{equation}
which can be solved numerically. 

The column density in the upper level of a $(J,K)$ inversion transition is given by
\begin{equation}
N_u(J,K) = \frac{3h}{8\pi^3}~\frac{1}{|\mu_{J,K}|^2}~\frac{1}{(e^{h\nu/kT_{\mathrm{ex}}}-1)}\int \tau(J,K)~dv
\end{equation}
where $v$ is the velocity. The total column density of a $(J,K)$ transition is related to the column density in the upper level by
\begin{equation}
N(J,K) = N_u(J,K)(1+e^{h\nu/kT_{\mathrm{ex}}})
\end{equation}
In addition, if the line shape is assumed to be a Gaussian, and using eq.~(\ref{tautranseq}), the integral of the line optical depth becomes
\begin{equation}
\int \tau(J,K)~dv = \frac{a\tau(J,K,m)\Delta v}{2\sqrt{(\ln 2)/\pi}}
\end{equation}
where $\Delta v$ is the full width at half maximum (FWHM) line width. Thus, the column density of the $(J,K)$ transition is given by
\begin{equation}
N(J,K) = 3.96 \times 10^{12} ~\frac{J(J+1)}{K^2} ~\frac{e^{h\nu/kT_{\mathrm{ex}}}+1}{e^{h\nu/kT_{\mathrm{ex}}}-1}~ a \tau(J,K,m) \Delta v~\mathrm{cm}^{-2}
\end{equation}
where $\Delta v$ is in km~s$^{-1}$. For the (1,1) transition, this becomes
\begin{equation}
N(1,1) = 1.58 \times 10^{13} ~\frac{e^{h\nu/kT_{\mathrm{ex}}}+1}{e^{h\nu/kT_{\mathrm{ex}}}-1} ~\tau(1,1,m) \Delta v~\mathrm{cm}^{-2}
\end{equation}
The total NH$_3$ column density can be estimated from $N(1,1)$ using
\begin{equation}
N(\mathrm{NH}_3) = N(1,1)\left(\frac{1}{3}~e^{23.3/T_{\mathrm{R21}}} + 1 + \frac{5}{3}~e^{-41.2/T_{\mathrm{R21}}} + \frac{14}{3}~e^{-100.3/T_{\mathrm{R21}}}\right)
\end{equation}

\end{appendix}

\begin{deluxetable}{lcccccccc}
\tabletypesize{\scriptsize}
\tablecaption{Properties of NH$_3$ lines towards the AMGPS sources.\label{table1}}
\tablewidth{0pt}
\tablehead{
\colhead{Source} & \colhead{$v_{\mathrm{LSR}}$} & \colhead{$T_{\mathrm{MB}}(1,1,m)$} & \colhead{$\Delta v(1,1)$} & \colhead{$\tau(1,1,m)$} & \colhead{$T_{\mathrm{MB}}(2,2,m)$} & \colhead{$\Delta v(2,2)$} & \colhead{$T_{\mathrm{MB}}(3,3,m)$} & \colhead{$\Delta v(3,3)$} \\
\colhead{} & \colhead{(km s$^{-1}$)} & \colhead{(K)} & \colhead{(km s$^{-1}$)} & \colhead{} & \colhead{(K)} & \colhead{(km s$^{-1}$)} & \colhead{(K)} & \colhead{(km s$^{-1}$)}
}
\startdata
G34.82$+$0.35  & 56.6 & 1.65 (0.03) & 2.35 (0.03) & 1.75 (0.06) & 0.90 (0.02) & 2.33 (0.07) & 0.22 (0.03) & 3.62 (0.25)  \\
G35.03$+$0.35  & 53.1 & 1.15 (0.07) & 3.15 (0.12) & 0.29 (0.16) & 0.75 (0.05) & 3.75 (0.15) & 0.53 (0.06) & 4.60 (0.21)  \\
G35.25$-$0.24  & 61.6 & 0.47 (0.03) & 1.43 (0.07) & 1.51 (0.27) & 0.16 (0.03) & 1.44 (0.19) & $< 0.09$    & \ldots       \\
G35.39$+$0.02  & 93.9 & 0.40 (0.03) & 1.30 (0.06) & 2.41 (0.34) & 0.16 (0.03) & 2.33 (0.51) & $< 0.09$    & \ldots       \\
G35.40$+$0.03  & 94.6 & 0.54 (0.03) & 1.86 (0.06) & 1.53 (0.20) & 0.29 (0.03) & 2.01 (0.13) & $< 0.09$    & \ldots       \\
G35.59$+$0.06  & 48.8 & 1.24 (0.05) & 2.35 (0.06) & 1.54 (0.13) & 0.64 (0.03) & 2.46 (0.09) & 0.26 (0.03) & 4.00 (0.21)  \\
G35.79$-$0.17  & 61.3 & 1.17 (0.09) & 2.40 (0.09) & 2.65 (0.27) & 0.67 (0.08) & 3.13 (0.18) & 0.57 (0.09) & 4.74 (0.46)  \\
G36.02$-$0.20  & 86.8 & 0.75 (0.05) & 1.55 (0.06) & 2.42 (0.28) & 0.31 (0.05) & 2.12 (0.25) & $< 0.15$    & \ldots       \\
G36.64$-$0.21  & 74.8 & 0.17 (0.04) & 0.73 (0.12) & 4.43 (1.70) & $< 0.09$    & \ldots      & $< 0.09$    & \ldots       \\
G36.70$+$0.09  & 59.4 & 0.56 (0.03) & 1.27 (0.05) & 1.60 (0.23) & 0.16 (0.03) & 1.56 (0.27) & $< 0.09$    & \ldots       \\
G36.84$-$0.02a & 56.8 & 0.53 (0.03) & 1.34 (0.09) & 3.49 (0.52) & 0.19 (0.03) & 1.58 (0.30) & \tablenotemark{\dag} &   \\
G36.84$-$0.02b & 59.0 & 0.72 (0.03) & 3.17 (0.18) & 1.54 (0.14) & 0.36 (0.03) & 4.63 (0.32) & \tablenotemark{\dag} &   \\
G36.90$-$0.41  & 79.6 & 1.29 (0.07) & 2.14 (0.57) & 2.07 (0.19) & 0.64 (0.05) & 2.07 (0.13) & 0.32 (0.06) & 2.99 (0.32)  \\
G36.92$+$0.48  & \ldots & $< 0.09$  & \ldots      & \ldots      & $< 0.09$    & \ldots      & $< 0.09$    & \ldots       \\
G37.02$-$0.03  & 80.1 & 1.14 (0.03) & 1.66 (0.02) & 2.31 (0.11) & 0.48 (0.03) & 2.06 (0.07) & 0.18 (0.03) & 2.55 (0.32)  \\
G37.04$-$0.04  & 80.8 & 1.70 (0.03) & 1.84 (0.01) & 2.23 (0.06) & 0.76 (0.03) & 2.35 (0.07) & 0.27 (0.03) & 3.88 (0.19)  \\
G37.38$-$0.09  & 57.1 & 0.15 (0.03) & 2.15 (0.39) & \ldots      & 0.11 (0.02) & 3.65 (0.49) & $< 0.08$    & \ldots       \\
G37.47$-$0.11  & 58.4 & 0.41 (0.03) & 2.45 (0.12) & 0.57 (0.22) & 0.24 (0.03) & 5.18 (0.50) & $< 0.09$    & \ldots       \\
G37.53$-$0.11  & 51.7 & 0.37 (0.07) & 2.43 (0.34) & \ldots      & 0.35 (0.07) & 3.72 (0.04) & 0.27 (0.06) & 2.86 (0.37)  \\
G37.55$+$0.19  & 84.6 & 1.95 (0.09) & 2.00 (0.05) & 2.47 (0.17) & 1.30 (0.08) & 2.61 (0.11) & 0.65 (0.08) & 3.24 (0.22)  \\
G37.60$+$0.42  & 89.2 & 0.78 (0.04) & 1.62 (0.05) & 1.96 (0.19) & 0.32 (0.03) & 3.04 (0.28) & 0.20 (0.03) & 2.38 (0.33)  \\
G37.74$-$0.12  & 45.5 & 0.67 (0.03) & 2.44 (0.08) & 1.14 (0.16) & 0.39 (0.03) & 3.31 (0.16) & 0.21 (0.03) & 5.20 (0.29)  \\
G37.76$-$0.19  & 59.6 & 0.47 (0.03) & 4.32 (0.07) & 1.05 (0.18) & 0.21 (0.03) & 2.85 (0.29) & 0.13 (0.03) & 2.08 (0.45)  \\
G37.77$-$0.22  & 64.1 & 1.04 (0.03) & 4.00 (0.04) & 1.37 (0.07) & 0.56 (0.03) & 4.90 (0.13) & 0.32 (0.03) & 5.34 (0.19)  \\
G38.03$-$0.30  & 61.6 & 0.70 (0.03) & 1.99 (0.04) & 1.04 (0.12) & 0.33 (0.03) & 2.24 (0.14) & 0.14 (0.03) & 2.71 (0.35)  \\
G38.08$-$0.27  & 64.7 & 0.27 (0.04) & 1.95 (0.22) & \ldots      & 0.09 (0.03) & 3.13 (0.57) & 0.11 (0.03) & 1.26 (0.30)  \\
G38.12$-$0.24  & 82.7 & 0.98 (0.03) & 1.77 (0.04) & 1.13 (0.12) & 0.49 (0.03) & 2.46 (0.12) & 0.16 (0.03) & 3.73 (0.48)  \\
G38.20$-$0.08  & 82.9 & 1.39 (0.07) & 2.75 (0.07) & 1.78 (0.15) & 0.54 (0.08) & 3.50 (0.29) & 0.33 (0.07) & 4.90 (0.42)  \\
G38.26$-$0.08  & \ldots & $< 0.09$  & \ldots      & \ldots      & $< 0.09$    & \ldots      & $< 0.09$    & \ldots       \\
G38.26$-$0.20  & 65.4 & 0.42 (0.03) & 1.80 (0.11) & 0.47 (0.28) & 0.16 (0.03) & 3.04 (0.39) & $< 0.09$    & \ldots       \\
G38.56$+$0.15  & 28.2 & 0.12 (0.03) & 2.51 (0.46) & \ldots      & $< 0.09$    & \ldots      & $< 0.09$    & \ldots       \\
G38.60$-$0.21  & 66.1 & 0.72 (0.04) & 2.05 (0.06) & 1.09 (0.18) & 0.24 (0.03) & 2.82 (0.22) & 0.13 (0.03) & 2.91 (0.38)  \\
G38.66$+$0.08  & $-$39.1 & 0.11 (0.02) & 2.85 (0.36) & 1.39 (0.65) & $< 0.09$ & \ldots      & $< 0.09$    & \ldots       \\
G38.92$-$0.36  & 37.9 & 2.81 (0.08) & 2.26 (0.05) & 1.13 (0.12) & 1.42 (0.08) & 2.54 (0.10) & 0.55 (0.08) & 4.02 (0.33)  \\
G39.39$-$0.14  & 66.0 & 0.72 (0.03) & 2.41 (0.08) & 0.63 (0.15) & 0.36 (0.03) & 2.91 (0.14) & 0.16 (0.03) & 4.72 (0.36)  \\
G39.54$-$0.38  & 60.4 & 0.39 (0.03) & 2.66 (0.12) & 0.82 (0.24) & 0.19 (0.03) & 2.14 (0.27) & $< 0.09$    & \ldots       \\
G40.28$-$0.22  & 73.1 & 1.59 (0.08) & 3.62 (0.10) & 0.54 (0.14) & 1.03 (0.10) & 3.88 (0.20) & 0.73 (0.07) & 5.58 (0.27)  \\
G40.62$-$0.14  & 32.5 & 1.00 (0.06) & 4.09 (0.10) & 1.15 (0.15) & 0.54 (0.06) & 4.27 (0.25) & 0.32 (0.06) & 5.50 (0.52)  \\
G40.94$-$0.04  & 39.3 & 0.12 (0.03) & 2.96 (0.52) & \ldots      & 0.09 (0.03) & 1.28 (0.30) & $< 0.09$    & \ldots       \\
G41.08$-$0.13  & 63.3 & 0.54 (0.03) & 2.79 (0.11) & 0.70 (0.17) & 0.22 (0.03) & 3.24 (0.27) & $< 0.09$    & \ldots       \\
G41.12$-$0.11  & 37.8 & 0.35 (0.03) & 1.45 (0.09) & 1.35 (0.36) & $< 0.09$    & \ldots      & $< 0.09$    & \ldots       \\
G41.12$-$0.22  & 59.7 & 0.52 (0.03) & 2.46 (0.08) & 1.03 (0.18) & 0.18 (0.03) & 2.25 (0.21) & 0.13 (0.03) & 3.66 (0.40)  \\
G41.16$-$0.20  & 60.4 & 0.35 (0.03) & 1.61 (0.13) & 1.56 (0.38) & 0.17 (0.03) & 2.10 (0.21) & $< 0.09$    & \ldots       \\
G41.23$-$0.20  & 58.9 & 0.50 (0.03) & 1.61 (0.06) & 1.36 (0.23) & 0.28 (0.03) & 2.03 (0.14) & $< 0.09$    & \ldots       \\
G41.27$+$0.37  & 14.9 & 0.26 (0.03) & 1.69 (0.14) & 2.58 (0.55) & 0.12 (0.02) & 2.08 (0.28) & $< 0.09$    & \ldots       \\
G41.34$-$0.14  & 13.2 & 0.43 (0.03) & 1.68 (0.07) & 2.46 (0.30) & 0.16 (0.03) & 2.11 (0.29) & $< 0.09$    & \ldots       \\
G41.58$+$0.04  & \ldots & $< 0.09$  & \ldots      & \ldots      & $< 0.09$    & \ldots      & $< 0.09$    & \ldots       \\
G42.03$+$0.19  & 17.7 & 0.19 (0.04) & 2.49 (0.36) & \ldots      & 0.10 (0.03) & 3.54 (0.86) & $< 0.09$    & \ldots       \\
G42.30$-$0.30  & 27.8 & 0.61 (0.03) & 2.58 (0.07) & 1.00 (0.14) & 0.22 (0.03) & 3.26 (0.24) & $< 0.09$    & \ldots       \\
G42.43$-$0.26  & 64.7 & 0.26 (0.04) & 2.28 (0.29) & \ldots      & 0.19 (0.03) & 2.83 (0.28) & 0.11 (0.03) & 3.12 (0.54)  \\
G42.70$-$0.15  & $-$44.4 & 0.11 (0.03) & 3.60 (0.78) & \ldots   & $< 0.09$    & \ldots      & $< 0.09$    & \ldots       \\
G43.04$-$0.46  & 57.4 & 0.69 (0.07) & 3.04 (0.12) & 2.01 (0.29) & 0.39 (0.06) & 3.83 (0.29) & 0.40 (0.07) & 4.30 (0.43)  \\
G43.08$-$0.08  & 12.6 & 0.52 (0.04) & 3.19 (0.14) & 0.45 (0.22) & 0.17 (0.04) & 2.86 (0.35) & $< 0.12$    & \ldots       \\
G44.31$+$0.04  & 56.2 & 1.44 (0.07) & 2.55 (0.06) & 1.24 (0.14) & 0.84 (0.07) & 2.70 (0.15) & 0.86 (0.07) & 2.82 (0.15)  \\
G44.64$-$0.52  & \ldots & $< 0.09$  & \ldots      & \ldots      & $< 0.09$    & \ldots      & $< 0.09$    & \ldots       \\
G45.44$+$0.07a & 54.8 & 0.22 (0.03) & 1.29 (0.14) & 1.08 (0.66) & 0.21 (0.03) & 2.04 (0.30) & $< 0.08$    & \ldots       \\
G45.44$+$0.07b & 59.0 & 0.38 (0.03) & 1.53 (0.11) & 0.75 (0.36) & 0.18 (0.03) & 2.97 (0.44) & 0.10 (0.03) & 4.54 (0.58)  \\
G45.47$+$0.13a & 57.1 & 0.35 (0.06) & 2.28 (0.24) & 1.01 (0.64) & 0.29 (0.07) & 2.19 (0.36) & $< 0.18$    & \ldots       \\
G45.47$+$0.13b & 61.6 & 0.84 (0.06) & 3.23 (0.13) & 0.96 (0.20) & 0.55 (0.07) & 4.34 (0.29) & 0.32 (0.06) & 3.71 (0.33)  \\
G45.47$+$0.05  & 60.8 & 0.64 (0.04) & 2.51 (0.09) & 1.30 (0.22) & 0.47 (0.04) & 3.34 (0.18) & 0.43 (0.04) & 3.18 (0.18)  \\
G45.49$+$0.13a & 59.5 & 1.49 (0.04) & 1.81 (0.06) & 1.56 (0.10) & 0.43 (0.04) & 1.44 (0.13) & \tablenotemark{\dag} &   \\
G45.49$+$0.13b & 62.2 & 1.13 (0.04) & 2.61 (0.12) & 1.08 (0.11) & 0.55 (0.04) & 5.00 (0.22) & \tablenotemark{\dag} &   \\
G45.57$-$0.12  &  4.5 & 0.21 (0.03) & 1.68 (0.15) & 2.06 (0.70) & $< 0.09$    & \ldots      & $< 0.09$    & \ldots       \\
G45.81$-$0.36  & 58.7 & 0.69 (0.04) & 3.05 (0.10) & 0.50 (0.16) & 0.37 (0.04) & 3.87 (0.22) & 0.22 (0.04) & 3.84 (0.34)  \\
G46.07$+$0.22  & 18.8 & 0.20 (0.03) & 1.69 (0.24) & \ldots      & $< 0.09$    & \ldots      & $< 0.09$    & \ldots       \\
G46.12$+$0.38  & 55.0 & 0.54 (0.03) & 1.72 (0.06) & 1.52 (0.23) & 0.17 (0.03) & 2.06 (0.26) & $< 0.09$    & \ldots       \\
G48.89$-$0.17  & 55.7 & 0.19 (0.02) & 1.34 (0.13) & 1.82 (0.53) & $< 0.07$    & \ldots      & $< 0.07$    & \ldots       \\
G48.90$-$0.27  & 68.4 & 1.60 (0.07) & 1.60 (0.05) & 1.46 (0.17) & 0.80 (0.06) & 2.23 (0.14) & 0.27 (0.06) & 3.68 (0.45)  \\
G48.99$-$0.30  & 67.4 & 2.48 (0.06) & 2.93 (0.04) & 1.40 (0.07) & 1.79 (0.06) & 3.31 (0.05) & 2.93 (0.08) & 3.14 (0.06)  \\
G49.27$+$0.31  &  3.3 & 0.59 (0.03) & 2.61 (0.08) & 1.27 (0.18) & 0.29 (0.03) & 3.87 (0.25) & 0.19 (0.03) & 4.89 (0.38)  \\
G49.35$+$0.41  & 66.2 & 0.26 (0.03) & 1.09 (0.10) & 1.83 (0.55) & $< 0.09$    & \ldots      & $< 0.09$    & \ldots       \\
G49.41$+$0.33  & $-$21.3 & 0.24 (0.03) & 2.30 (0.25) & \ldots   & 0.10 (0.03) & 2.89 (0.50) & $< 0.09$    & \ldots       \\
G49.60$-$0.25  & 56.7 & 1.55 (0.03) & 1.72 (0.03) & 1.18 (0.08) & 0.74 (0.03) & 2.00 (0.06) & 0.22 (0.03) & 2.46 (0.20)  \\
G49.62$-$0.36  & 54.5 & 0.21 (0.03) & 1.76 (0.22) & \ldots      & $< 0.08$    & \ldots      & $< 0.08$    & \ldots       \\
G50.78$+$0.15  & 42.1 & 0.83 (0.03) & 1.84 (0.04) & 1.80 (0.13) & 0.41 (0.03) & 2.45 (0.12) & 0.11 (0.03) & 2.86 (0.35)  \\
G52.92$+$0.41  & 44.7 & 1.32 (0.03) & 1.65 (0.02) & 1.33 (0.07) & 0.52 (0.04) & 2.12 (0.11) & 0.13 (0.03) & 4.88 (0.48)  \\
G53.04$+$0.11  &  4.8 & 0.88 (0.04) & 2.61 (0.07) & 0.58 (0.13) & 0.40 (0.03) & 3.39 (0.15) & 0.19 (0.04) & 3.35 (0.36)  \\
G53.14$+$0.07  & 21.7 & 1.69 (0.05) & 1.89 (0.04) & 0.89 (0.11) & 0.91 (0.06) & 2.31 (0.12) & 0.32 (0.05) & 3.73 (0.37)  \\
G53.62$+$0.04  & 22.8 & 1.35 (0.03) & 1.31 (0.02) & 1.30 (0.09) & 0.57 (0.03) & 1.62 (0.06) & 0.13 (0.03) & 2.78 (0.42)  \\
\enddata
\tablecomments{The columns show the NH$_3$ (1,1), (2,2) and (3,3) line fit parameters and their uncertainties returned by the CLASS software. Sources for which no $\tau(1,1,m)$ is listed had their (1,1) spectra fit using Gaussians due to the line being too weak. Note that the uncertainties in $T_{\mathrm{MB}}$ only include the formal fitting errors and do not include the uncertainty in flux calibration. No uncertainties are shown for the line velocities since they are well below 0.1~km~s$^{-1}$. All the upper limits quoted are $3\sigma$. The sources G43.80$-$0.13 and G45.07$+$0.13 are not included in the Table due to poor fits to their complex line profiles and blending from multiple velocity components.}
\tablenotetext{\dag}{No reliable fit possible due to blending from two closely separated velocity components.}
\end{deluxetable}

\begin{deluxetable}{lccccc}
\tabletypesize{\footnotesize}
\tablecaption{Physical parameters determined from the NH$_3$ observations.\label{table2}}
\tablewidth{0pt}
\tablehead{
\colhead{Source} & \colhead{$T_{\mathrm{ex}}(1,1)$} & \colhead{$T_{\mathrm{R21}}$} & \colhead{$T_\mathrm{k}$} & \colhead{$T_{\mathrm{R31}}$} & \colhead{$N$(NH$_3$)} \\
\colhead{} & \colhead{(K)} & \colhead{(K)} & \colhead{(K)} & \colhead{(K)} & \colhead{($10^{15}$ cm$^{-2}$)} 
}
\startdata
G34.82$+$0.35  & 4.7 (0.2) & 17.6 (1.5) & 23.4 (3.2) & 26.1 (1.4) & 4.9 (0.5)  \\
G35.03$+$0.35  & 7.4 (2.3) & 23.6 (2.5) & 40.0 (9.1) & 21.7 (1.2) & 1.3 (0.8)  \\
G35.25$-$0.24  & 3.3 (0.1) & 14.4 (1.5) & 17.8 (2.7) & \ldots     & 2.4 (0.5)  \\
G35.39$+$0.02  & 3.2 (0.1) & 13.8 (1.5) & 16.7 (2.6) & \ldots     & 3.5 (0.7)  \\
G35.40$+$0.03  & 3.4 (0.1) & 18.1 (2.0) & 24.5 (4.4) & \ldots     & 3.4 (0.6)  \\
G35.59$+$0.06  & 4.3 (0.2) & 17.6 (1.6) & 23.5 (3.4) & 19.2 (0.9) & 4.3 (0.6)  \\
G35.79$-$0.17  & 4.0 (0.2) & 16.4 (2.0) & 21.0 (3.9) & 20.8 (1.0) & 7.3 (1.3)  \\
G36.02$-$0.20  & 3.6 (0.1) & 14.3 (1.5) & 17.6 (2.6) & \ldots     & 4.2 (0.7)  \\
G36.70$+$0.09  & 3.4 (0.1) & 13.5 (1.3) & 16.2 (2.1) & \ldots     & 2.3 (0.4)  \\
G36.84$-$0.02a & 3.3 (0.1) & 12.4 (1.2) & 14.3 (1.9) & \ldots     & 5.2 (1.2)  \\
G36.84$-$0.02b & 3.7 (0.1) & 17.3 (1.7) & 22.8 (3.5) & \ldots     & 5.8 (0.8)  \\
G36.90$-$0.41  & 4.2 (0.2) & 16.1 (1.5) & 20.4 (2.9) & 26.3 (1.5) & 5.1 (0.7)  \\
G37.02$-$0.03  & 4.0 (0.1) & 14.5 (1.0) & 17.9 (1.9) & 27.6 (2.1) & 4.3 (0.4)  \\
G37.04$-$0.04  & 4.7 (0.2) & 15.0 (1.1) & 18.9 (2.0) & 19.9 (0.9) & 4.6 (0.4)  \\
G37.38$-$0.09  & 4.1\tablenotemark{\dag} (0.9) & 26.9 (6.4) & 48.4 ($^{+31.3}_{-17.8}$) & \ldots & 4.8 (0.4)  \\
G37.47$-$0.11  & 3.7 (0.3) & 21.2 (2.6) & 32.2 (7.4) & \ldots     & 1.9 (0.8)  \\
G37.53$-$0.11  & 4.1\tablenotemark{\dag} (0.9) & 31.4 (9.1) & 67.3 ($^{+93.6}_{-31.9}$) & 30.5 (3.0) & 1.8 (1.6)  \\
G37.55$+$0.19  & 4.9 (0.2) & 18.5 (2.2) & 25.4 (5.1) & 17.4 (0.8) & 6.0 (0.9)  \\
G37.60$+$0.42  & 3.6 (0.1) & 14.9 (1.3) & 18.7 (2.4) & 26.0 (1.6) & 3.6 (0.5)  \\
G37.74$-$0.12  & 3.7 (0.1) & 19.7 (2.1) & 28.2 (5.4) & 38.5 (4.5) & 3.5 (0.6)  \\
G37.76$-$0.19  & 3.5 (0.1) & 17.2 (1.9) & 22.6 (4.0) & 40.1 (4.6) & 2.9 (0.6)  \\
G37.77$-$0.22  & 4.1 (0.1) & 18.4 (1.7) & 25.1 (3.8) & 32.6 (2.4) & 6.7 (0.7)  \\
G38.03$-$0.30  & 3.8 (0.1) & 17.7 (1.6) & 23.7 (3.5) & 25.4 (1.4) & 2.5 (0.4)  \\
G38.12$-$0.24  & 4.2 (0.2) & 18.1 (1.6) & 24.5 (3.5) & 55.2 (18.8) & 2.4 (0.3)  \\
G38.20$-$0.08  & 4.4 (0.2) & 14.8 (1.4) & 18.6 (2.6) & 21.6 (1.0) & 5.5 (0.8)  \\
G38.26$-$0.20  & 3.9 (0.5) & 17.1 (2.0) & 22.4 (4.1) & \ldots     & 1.0 (0.6)  \\
G38.60$-$0.21  & 3.8 (0.2) & 15.1 (1.3) & 19.1 (2.5) & 29.2 (1.9) & 2.5 (0.5)  \\
G38.92$-$0.36  & 6.9 (0.5) & 18.3 (1.6) & 24.8 (3.6) & 17.9 (0.7) & 3.1 (0.4)  \\
G39.39$-$0.14  & 4.3 (0.3) & 19.5 (1.8) & 27.7 (4.4) & 28.8 (1.7) & 1.9 (0.5)  \\
G39.54$-$0.38  & 3.4 (0.2) & 18.5 (2.3) & 25.4 (5.4) & \ldots     & 2.7 (0.9)  \\
G40.28$-$0.22  & 6.6 (0.9) & 22.9 (2.7) & 37.7 (9.0) & 25.7 (1.2) & 2.8 (0.8)  \\
G40.62$-$0.14  & 4.2 (0.2) & 18.9 (2.2) & 26.4 (5.1) & 29.2 (2.3) & 5.8 (1.0)  \\
G41.08$-$0.13  & 3.8 (0.2) & 17.3 (1.8) & 22.8 (3.7) & \ldots     & 2.3 (0.6)  \\
G41.12$-$0.22  & 3.6 (0.1) & 15.4 (1.6) & 19.7 (3.1) & 24.5 (1.7) & 2.9 (0.6)  \\
G41.16$-$0.20  & 3.2 (0.1) & 16.9 (2.2) & 22.0 (4.5) & \ldots     & 2.9 (0.9)  \\
G41.23$-$0.20  & 3.4 (0.1) & 18.9 (2.2) & 26.2 (5.2) & \ldots     & 2.7 (0.6)  \\
G41.27$+$0.37  & 3.0 (0.1) & 14.8 (2.1) & 18.4 (3.9) & \ldots     & 4.9 (1.4)  \\
G41.34$-$0.14  & 3.2 (0.1) & 13.7 (1.5) & 16.6 (2.6) & \ldots     & 4.6 (0.9)  \\
G42.30$-$0.30  & 3.7 (0.1) & 15.7 (1.4) & 20.3 (2.9) & \ldots     & 2.9 (0.5)  \\
G42.43$-$0.26  & 4.1\tablenotemark{\dag} (0.9) & 25.4 (4.5) & 42.5 (15.6) & \ldots & 0.9 (0.7)  \\
G43.04$-$0.46  & 3.5 (0.1) & 17.7 (2.6) & 23.5 (5.7) & 23.7 (1.6) & 7.3 (1.6)  \\
G43.08$-$0.08  & 4.2 (0.6) & 16.2 (1.9) & 20.6 (3.7) & \ldots     & 1.7 (0.8)  \\
G44.31$+$0.04  & 4.8 (0.3) & 19.7 (2.2) & 28.3 (5.5) & 39.4 (4.8) & 4.0 (0.7)  \\
G45.44$+$0.07a & 3.1 (0.1) & 29.7 (8.9) & 61.3 ($^{+68.2}_{-29.9}$) & \ldots & 2.7 (1.8)  \\
G45.44$+$0.07b & 3.5 (0.3) & 18.6 (2.4) & 25.7 (5.6) & \ldots     & 1.4 (0.7)  \\
G45.47$+$0.13a & 3.3 (0.2) & 26.3 (8.4) & 45.9 ($^{+49.2}_{-21.9}$) & \ldots & 3.8 (2.7)  \\
G45.47$+$0.13b & 4.1 (0.2) & 22.0 (3.2) & 34.6 (9.8) & 27.4 (1.6) & 4.3 (1.1)  \\
G45.47$+$0.05  & 3.6 (0.1) & 22.9 (3.5) & 37.5 (11.7) & 23.7 (1.5) & 4.7 (1.1)  \\
G45.49$+$0.13a & 4.6 (0.2) & 13.5 (0.9) & 16.2 (1.5) & \ldots     & 3.1 (0.4)  \\
G45.49$+$0.13b & 4.5 (0.2) & 18.0 (1.6) & 24.3 (3.5) & \ldots     & 3.4 (0.5)  \\
G45.81$-$0.36  & 4.5 (0.5) & 20.5 (2.2) & 30.5 (6.1) & 51.3 (7.7) & 2.0 (0.7)  \\
G46.12$+$0.38  & 3.4 (0.1) & 14.0 (1.3) & 17.0 (2.4) & \ldots     & 2.9 (0.6)  \\
G48.90$-$0.27  & 4.8 (0.3) & 17.4 (1.6) & 22.9 (3.4) & 24.4 (1.4) & 2.8 (0.4)  \\
G48.99$-$0.30  & 6.0 (0.3) & 22.4 (2.7) & 35.8 (8.6) & 18.1 (0.7) & 5.8 (0.7)  \\
G49.27$+$0.31  & 3.6 (0.1) & 17.6 (1.8) & 23.3 (3.8) & 24.1 (1.6) & 3.9 (0.7)  \\
G49.60$-$0.25  & 5.0 (0.2) & 17.6 (1.4) & 23.3 (2.9) & 19.2 (0.8) & 2.4 (0.3)  \\
G50.78$+$0.15  & 3.7 (0.1) & 16.6 (1.5) & 21.4 (2.9) & 23.8 (1.2) & 3.8 (0.5)  \\
G52.92$+$0.41  & 4.5 (0.2) & 15.7 (1.1) & 20.3 (2.3) & 18.4 (0.9) & 2.5 (0.2)  \\
G53.04$+$0.11  & 4.8 (0.4) & 18.7 (1.6) & 25.8 (3.8) & 22.6 (1.4) & 1.9 (0.5)  \\
G53.14$+$0.07  & 5.6 (0.4) & 19.5 (1.8) & 27.9 (4.6) & 20.3 (0.9) & 2.1 (0.3)  \\
G53.62$+$0.04  & 4.6 (0.2) & 16.3 (1.2) & 20.8 (2.3) & 23.8 (1.4) & 2.0 (0.2)  \\
\enddata
\tablecomments{The columns show the source name, excitation temperature, $T_{\mathrm{ex}}$, rotation temperation from (2,2) and (1,1), $T_{\mathrm{R21}}$, kinetic temperature, $T_\mathrm{k}$, rotation temperature from (3,3) and (1,1), $T_{\mathrm{R31}}$, and total NH$_3$ column density, N(NH$_3$). The uncertainties include contributions from the flux calibration. Only sources that have a $4\sigma$ detection or higher are shown in this table.}
\tablenotetext{\dag}{$T_{\mathrm{ex}}$ assumed to be $4.1 \pm 0.9$ K since the optical depth of the (1,1) main hyperfine component is not constrained.}
\end{deluxetable}

\begin{deluxetable}{lcccclccc}
\tabletypesize{\footnotesize}
\tablecaption{H$_2$ column densities and NH$_3$ abundances derived from the Bolocam Galactic Plane Survey.\label{table3}}
\tablewidth{0pt}
\tablehead{
\colhead{Source} & \colhead{$S_{\mathrm{1.1mm}}$} & \colhead{$N_{\mathrm{H}_2}$} & \colhead{$\chi_{\mathrm{NH}_3}$} & \colhead{} & \colhead{Source} & \colhead{$S_{\mathrm{1.1mm}}$} & \colhead{$N_{\mathrm{H}_2}$} & \colhead{$\chi_{\mathrm{NH}_3}$} \\
\colhead{} & \colhead{(Jy)} & \colhead{($10^{21}$ cm$^{-2}$)} & \colhead{($\times 10^{-7}$)} & \colhead{} & \colhead{} & \colhead{(Jy)} & \colhead{($10^{21}$ cm$^{-2}$)} & \colhead{($\times 10^{-7}$)} 
}
\startdata
34.82$+$0.35 & 1.02 (0.08) & 28.0 (5.5) & 1.7 (0.4) & & 39.39$-$0.14 & 0.53 (0.05) & 13.2 (2.9) & 1.5 (0.5) \\
35.03$+$0.35 & 1.38 (0.09) & 11.6 (3.2) & 1.2 (0.7) & & 40.28$-$0.22 & 1.57 (0.11) & 25.1 (7.3) & 1.1 (0.4) \\
35.25$-$0.24 & 0.09 (0.04) &  6.7 (2.9) & 3.6 (1.8) & & 40.62$-$0.14 & 0.83 (0.07) & 18.0 (4.7) & 3.2 (1.0) \\
35.40$+$0.03 & 0.19 (0.04) &  4.7 (1.3) & 7.4 (2.4) & & 41.08$-$0.13 & 0.25 (0.04) &  6.7 (1.7) & 3.4 (1.3) \\
35.79$-$0.17 & 0.75 (0.06) & 22.9 (6.0) & 3.2 (1.0) & & 41.12$-$0.22 & 0.30 (0.05) & 10.3 (2.5) & 2.8 (0.9) \\
36.02$-$0.20 & 0.24 (0.04) & 12.4 (3.1) & 3.4 (1.0) & & 41.27$+$0.37 & 0.14 (0.04) &  7.0 (2.5) & 7.0 (3.2) \\
36.70$+$0.09 & 0.16 (0.04) &  8.6 (2.0) & 2.6 (0.8) & & 41.34$-$0.14 & 0.13 (0.04) &  8.9 (3.1) & 5.1 (2.0) \\
36.84$-$0.02 & 0.37 (0.05) & 20.9 (4.7) & 5.3 (1.4) & & 42.30$-$0.30 & 0.32 (0.05) & 12.6 (2.9) & 2.3 (0.7) \\
36.90$-$0.41 & 0.26 (0.04) &  8.9 (2.0) & 5.7 (1.5) & & 42.43$-$0.26 & 0.43 (0.05) &  6.5 (2.8) & 1.4 (1.2) \\
37.04$-$0.04 & 0.46 (0.05) & 12.0 (2.1) & 3.8 (0.7) & & 43.04$-$0.46 & 0.79 (0.07) & 26.8 (8.9) & 2.7 (1.1) \\
37.38$-$0.09 & 0.15 (0.04) &  2.5 ($^{+1.9}_{-1.2}$) & 1.9 ($^{+1.8}_{-2.1}$) & & 43.08$-$0.08 & 0.24 (0.06) & 10.3 (3.3) & 1.6 (1.0) \\
37.47$-$0.11 & 0.30 (0.06) &  6.8 (2.2) & 2.8 (1.4) & & 44.31$+$0.04 & 0.76 (0.07) & 15.1 (3.9) & 2.7 (0.8) \\
37.53$-$0.11 & 0.74 (0.06) &  5.1 ($^{+7.8}_{-2.7}$) & 3.6 ($^{+3.7}_{-6.3}$) & & 45.47$+$0.05 & 1.30 (0.11) & 19.5 (7.3) & 2.4 (1.1) \\
37.55$+$0.19 & 1.04 (0.09) & 24.5 (6.6) & 2.5 (0.8) & & 45.81$-$0.36 & 0.42 (0.05) & 11.9 (3.2) & 1.7 (0.7) \\
37.60$+$0.42 & 0.30 (0.04) & 17.0 (3.8) & 2.1 (0.6) & & 48.99$-$0.30 & 2.00 (0.14) & 27.6 (8.1) & 2.1 (0.7) \\
37.74$-$0.12 & 0.75 (0.06) & 12.1 (3.1) & 2.9 (0.9) & & 49.27$+$0.31 & 0.25 (0.05) & 10.2 (2.9) & 3.9 (1.3) \\
37.76$-$0.19 & 0.35 (0.06) & 12.0 (3.3) & 2.4 (0.8) & & 49.60$-$0.25 & 0.36 (0.06) & 14.5 (3.3) & 1.7 (0.4) \\
37.77$-$0.22 & 1.26 (0.09) & 35.7 (7.4) & 1.9 (0.4) & & 50.78$+$0.15 & 0.27 (0.06) & 11.7 (2.9) & 3.3 (0.9) \\
38.03$-$0.30 & 0.18 (0.04) &  7.5 (2.1) & 3.3 (1.0) & & 52.92$+$0.41 & 0.36 (0.07) & 16.1 (3.6) & 1.5 (0.4) \\
38.12$-$0.24 & 0.27 (0.05) &  9.5 (2.3) & 2.5 (0.7) & & 53.04$+$0.11 & 0.56 (0.08) & 14.6 (3.4) & 1.3 (0.4) \\
38.20$-$0.08 & 0.44 (0.05) & 18.6 (4.2) & 2.9 (0.8) & & 53.14$+$0.07 & 1.48 (0.12) & 30.4 (6.7) & 0.7 (0.2) \\
38.60$-$0.21 & 0.20 (0.04) &  8.5 (2.0) & 3.0 (0.9) & & 53.62$+$0.04 & 0.62 (0.09) & 23.9 (4.7) & 0.8 (0.2) \\
\enddata
\tablecomments{The columns show the source name, 1.1~mm integrated flux from BGPS, H$_2$ column density and NH$_3$ abundance. To calculate the H$_2$ column densities, all BGPS fluxes listed above have been multiplied by a factor of 1.5 to conform with measurements of previous surveys as indicated in \citet{agui11}.}
\end{deluxetable}

\clearpage

\end{document}